\documentclass{ifacconf}

\usepackage{graphicx}      
\usepackage{natbib}        

\newcommand{\epsc}{\epsilon_\mathrm{c}}
\newcommand{\epsd}{\epsilon_\mathrm{d}}

\usepackage{amsmath}
\usepackage{amssymb}
\usepackage{bm}
\usepackage{makecell}
\usepackage{algorithm}
\usepackage{algpseudocode}
\usepackage{multirow}
\usepackage{color}
\usepackage{pgfplots}
\pgfplotsset{compat=newest}
\usepackage{url}

\begin{document}
\begin{frontmatter}

\title{A state reduction approach for learning-based model predictive control for train rescheduling} 

\author[First]{C.F.O da Silva \thanksref{equalcontribution}} 
\author[First]{X. Liu \thanksref{equalcontribution}} 
\author[First]{A. Dabiri}
\author[First]{B. De Schutter}

\thanks[equalcontribution]{These authors contributed equally to this work.}

\address[First]{Delft University of Technology, The Netherlands (e-mail: \{c.f.oliveiradasilva, x.liu-20, a.dabiri, b.deschutter\}@tudelft.nl).}

\begin{abstract}                
This paper proposes a state reduction method for learning-based model predictive control (MPC) for train rescheduling in urban rail transit systems.
The state reduction integrates into a control framework where the discrete decision variables are determined by a learning-based classifier and the continuous decision variables are computed by MPC.
Herein, the state representation is designed separately for each component of the control framework. While a reduced state is employed for learning, a full state is used in MPC.
Simulations on a large-scale train network highlight the effectiveness of the state reduction mechanism in improving the performance and reducing the memory usage.
\end{abstract}


\end{frontmatter}


\section{Introduction}
Urban rail transit systems play an important role in modern metropolitan mobility due to their high efficiency, capacity, and environmental sustainability. With the growing demand for public transportation and the rapid expansion of urban rail transit networks, ensuring the safety, punctuality, and adaptability of train operations has become increasingly important. Real-time train scheduling improves passenger satisfaction by reducing waiting and transfer times, and helps minimize operational costs under infrastructure and resource restrictions. 
Therefore, effective real-time train rescheduling approaches are essential to maintain service quality, particularly during disruptions, peak-hour congestion, and unexpected events. 

Formulating train scheduling problems typically leads to a mixed-integer programming problem with operational constraints.
Model predictive control (MPC) has emerged as an effective framework for real-time train scheduling, due to its ability to incorporate control and state constraints explicitly. \cite{de2002model} applied MPC to train scheduling by formulating a framework that dynamically adjusts transfer connections to minimize overall train delays.  \cite{caimi2012model} proposed an MPC framework for coordinating transfers and train allocations of timetable optimization in complex station environments. \cite{liu2023modeling} developed a simplified passenger flow model by approximating time-varying passenger demands as piecewise constant over fixed intervals, and then a mixed-integer linear programming (MILP)-based MPC approach is introduced for real-time train rescheduling. MPC has also been introduced for real-time train scheduling during various scenarios, including disturbances \citep{wang2020event}, disruptions \citep{cavone2020mpc}, and uncertain passenger flows \citep{liu2024real}.
However, MPC encounters computational challenges in solving large-scale mixed-integer programming problems when applied to real-time train scheduling.

To address the growing complexity and uncertainty in urban rail operations, learning-based approaches 
have gained increasing attention in the research of train scheduling problems. 
\cite{gattermann2023using} introduced a random forest classifier to replace complex penalty structures in crew scheduling by predicting planners’ preferences, achieving higher acceptance rates with minimal cost increase. 
\cite{ying2022adaptive} applied proximal policy optimization to train scheduling with flexible train composition, using neural networks for policy learning and a masking scheme to enforce constraint satisfaction. \cite{kuppusamy2020deep} addressed energy-efficient timetable rescheduling by training a long short-term memory (LSTM) network to select optimal train operation modes under varying conditions. \cite{wang2022shortening} designed a deep neural network to optimize train dwell times in metro systems by balancing the waiting time on the platforms and the travel time onboard. By leveraging gated recurrent units and graph attention networks, it captures spatio-temporal passenger flow patterns and inter-train interactions. However, learning-based approaches typically have constraint satisfaction issues, which limit their applicability, as safety remains a crucial requirement in railway operations.

Recent studies on mixed-integer programming problems have explored the integration of learning-based and optimization-based approaches, which aim to combine the computational efficiency of learning methods with the constraint satisfaction capabilities of optimization-based approaches.
In \cite{cauligi22_PRISM} and \cite{masti2020}, the discrete and continuous decision variables of the mixed-integer programming problem are determined separately by a supervised learning classifier and by the solution of an optimization problem, respectively. In a similar setting, \cite{dasilva2024integrating} applied reinforcement to determine the discrete decision variables and optimization to compute the continuous decision variables.
Integrated approaches have also been explored in train scheduling problems. \cite{zhang2024integrated} combined reinforcement learning with constrained optimization for train rescheduling, where integer variables were determined using a double deep Q-network, after which the MILP problem was transformed into a linear programming problem for efficient solution. However, the proposed learning-based approach has scaling issues and is limited to small-scale cases. 
Our recent work \citep{liu2025learningbasedmodelpredictivecontrol} integrated machine learning into MPC to better capture passenger flow patterns and improve rescheduling responsiveness. To enhance computational efficiency, presolve techniques were developed to reduce the integer solution space, and an LSTM network was trained to predict integer variables. Nonetheless, the use of learning introduced a modest degradation in closed-loop performance.

Based on previous work in \cite{liu2025learningbasedmodelpredictivecontrol}, the current paper introduces a state reduction approach for learning-based MPC for train rescheduling.
Specifically, a multi-resolution state design is integrated into the control framework, where a low-resolution state space is utilized for learning, while the full state space is employed in MPC.
For the same case study as in \cite{liu2025learningbasedmodelpredictivecontrol}, we demonstrate that the proposed method improves closed-loop performance and reduces computational resource usage. 

The remainder of this paper is structured as follows. Section~\ref{problem} introduces the train scheduling model and the corresponding problem description. Section~\ref{method} introduces the proposed state reduction methodology and its integration with the learning-based MPC approach. Section~\ref{case_study} presents a case study to illustrate the effectiveness of the developed approach. Section~\ref{conclusion} concludes the paper and outlines future research directions.

\section{Problem description}\label{problem}

We apply the passenger flow model developed in \cite{liu2025learningbasedmodelpredictivecontrol} for the train rescheduling problem, where trains depart at regular intervals, time-varying passenger flows are approximated as piecewise constant functions over fixed time windows, and the composition of trains is flexibly adjusted to meet real-time passenger demand.  A concise overview of the model and its associated optimization problem is provided below. For comprehensive descriptions and theoretical foundations, readers are referred to \cite{liu2025learningbasedmodelpredictivecontrol}.

We consider the problem of minimizing passenger delays and operational costs within the operational constraints. In urban rail transit systems, a train service starts at the origin station, stops at each station along the route, and then either returns to the depot or links with another service at the depot.  The total passenger delay relative to platform $p$ and train service $k_p$ is defined as
\begin{align}\label{eq:cost_passenger_nonlinear}
J_p^\mathrm{pass}(k_p) = &n_{p}(k_p)\left( {d_{p}(k_p) - d^\mathrm{pre}_{p}(k_p)} \right)+ \nonumber\\
&+ n^\mathrm{after}_{p}(k_p)\left( { d^\mathrm{pre}_{p}(k_p+1) -  d_{p}(k_p)} \right)
\end{align}
where $d^\mathrm{pre}_p(k_p)$ represents the predetermined departure time of train service $k_p$ at platform $p$ in the original timetable, and $d_p(k_p)$ is the actual departure time of train service $k_p$ at platform $p$, $n_p(k_p)$ and $n_p^\mathrm{after}(k_p)$ are the number of passengers at platform $p$ at time $d^\mathrm{pre}_p(k_p)$ and immediately after $d_p(k_p)$, respectively. 
The operational costs are described by
\begin{equation}\label{lmpc-cost}
{J_p^\mathrm{cost}}(k_p) = \ell_{p}(k_p){E_p^\mathrm{energy}} + \eta_{k_p,p}{E_p^\mathrm{add}},
\end{equation}
where $\ell_{p}(k_p)$ is the train composition (i.e., the number of train units) of train service $k_p$ when it departs from platform $p$, $E_p^\mathrm{energy}$ is the average energy consumption for the train service from platform $p$ to the next platform, $\eta_{k_p,p}$ is a binary variable describing whether the train composition of the train service is changed, and $E_p^\mathrm{add}$ is the cost for changing the train composition. The cost \eqref{eq:cost_passenger_nonlinear} is nonlinear, but it can be approximated by a linear expression by using upper and lower bounds for the variable $d_p(k_p) \in [d^\mathrm{pre}_{p}(k_p), d^\mathrm{pre}_{p}(k_p+1)\big)$ as follows:
\begin{align}\label{eq:cost_passenger_linear}
J_p^\mathrm{pass}(k_p) \approx & \ w_3 n_{p}(k_p)\left( {d^\mathrm{pre}_{p}(k_p+1) - d^\mathrm{pre}_{p}(k_p)} \right)+ \nonumber\\
&+ n^\mathrm{after}_{p}(k_p)\left( { d^\mathrm{pre}_{p}(k_p+1) -  d^\mathrm{pre}_{p}(k_p)} \right),
\end{align}
where $w_3$ is a weighing factor used to reduce the approximation error. 

The passenger flow model and the corresponding optimization problem are given as:
\begin{subequations}\label{problem_original}
\small
\begin{align}
&\mathop {\min }   J({\kappa_0})\! :=\! \sum\limits_{p \in \mathcal{P}}\! \sum\limits_{\ k_p \!\in \!\mathcal{N}_p(\!\kappa_0)} \!\left( \!{w_1 J_p^\mathrm{pass}(k_p) + w_2 J_p^\mathrm{cost}(k_p)} \right),\\
&\mathrm{subject}\quad\mathrm{to:} \nonumber\\
 & d^\mathrm{pre}_{p}(k_p) \le d_{p}(k_p) < d^\mathrm{pre}_{p}(k_p+1),\\
  &d_{p}(k_p) = a_{p}(k_p) + \tau_{p}(k_p),\\
  &a_{p}(k_p+1) = d_{p}(k_p) + h_{p}(k_p),\\
  &h_{p}(k_p) \ge h^\mathrm{min}_{p},\\
& n_{p}(k_p+1) = n_{p}(k_p) + \rho_{p}(k_p\!+\!1)\left(d^\mathrm{pre}_{p}(k_p\!+\!1) - d^\mathrm{pre}_{p}(k_p) \right) \nonumber\\
& \qquad \qquad  \qquad + n^\mathrm{trans}_{p}(k_p) - n^\mathrm{depart}_{p}(k_p),  \\
& n^\mathrm{depart}_{p}(k_p) \le C_{p}(k_p),\\
& C_{p}(k_p) = \ell_{p}(k_p) C_\mathrm{max},\\
& \ell_{\min}\le \ell_{p}(k_{p}) \le \ell_{\max},\\
& n^\mathrm{depart}_{p}(k_p) \le n^\mathrm{before}_{p}(k_p) ,\\
& n^\mathrm{before}_{p}(k_p) = n_{p}(k_p) + \rho_{p}(k_p+1) \left(d_{p}(k_p) - d^\mathrm{pre}_{p}(k_p) \right) \nonumber\\
& \qquad \qquad  \qquad + n^\mathrm{trans}_{p}(k_p),\\
& n^\mathrm{arrive}_{p}(k_p) =  n^\mathrm{depart}_{\mathrm{p}^\mathrm{pla}\left( p \right)}(k_p),\\
& n^\mathrm{after}_{p}(k_p) = n^\mathrm{before}_{p}(k_p) - n^\mathrm{depart}_{p}(k_p).
\end{align}
\end{subequations}
where $a_p(k_p)$, $\tau_p(k_p)$, $h_p(k_p)$ represent the decision variables for the arrival time, dwell time, and headway of train service $k_p$ at platform $p$, respectively; $\rho_p(k_p+1)$ is the passenger flow for train service $k_p+1$ at platform $p$, $n^\mathrm{depart}_{p}(k_p)$ is the number of passengers that boarded train service $k_p$, and $n^\mathrm{trans}_{p}(k_p)$ is the number of passengers that transfer to platform $p$ since the departure of train service $k_p$; $C_{p}(k_p)$ demotes the total capacity of train service $k_p$, and $C_\mathrm{max}$ is the capacity of each train unit. The integer decision variables are the order of the trains at the terminal station of a line, and the number of train compositions $\ell_{p}(k_{p})$ when train service $k_{p}$ departs from the terminal station. 

By using the transformation method in \cite{bemporad1999control,williams2013model}, the finite-horizon optimal control problem for the MPC controller can finally be defined as  \citep{liu2025learningbasedmodelpredictivecontrol}:
\begin{subequations}\label{eq:minlp}
\begin{align}
&{\mathop {\min}\limits_{\scriptstyle\substack{{\bm{x}}(\kappa_0),\\ {\bm{\epsc}}(\kappa_0),{\bm{\epsd}}(\kappa_0)}}  J^{(\mathrm{MPC})}(\chi(\kappa_0)) := \sum\limits_{\kappa = {\kappa_0}}^{{\kappa_0}+ {N_p} - 1} {L(x(\kappa),\epsc(\kappa),{\epsd}(\kappa))} } \label{lmpc-obj}\\
&\quad {\rm{s}}.{\rm{t}}.\quad {x}(\kappa + 1) = {A_\kappa}{x}(\kappa) + {B_{1,\kappa}}{\epsc}(\kappa) + {B_{2,\kappa}}{\epsd}(\kappa),\label{eq:minlp_state_dynamics}\\
&\quad \qquad {D_{3,\kappa}}{x}(\kappa) + {D_{1,\kappa}}{\epsc}(\kappa) +  {D_{2,\kappa}}{\epsd}(\kappa)  \le  {D_{4,\kappa}},\label{eq:minlp_constraints}\\
&\quad  \qquad \kappa = {\kappa_0}, \cdots ,{\kappa_0} + N_p - 1, \nonumber
\end{align}
\end{subequations}
where $ J^{(\mathrm{MPC})}(\chi (\kappa_0))$ represents the cost function as defined in (\ref{problem_original}), ${\bm{\epsc}}(\kappa_0) = [{\epsc}^\intercal (\kappa_0), {\epsc}^\intercal(\kappa_0+1), \ldots, {\epsc}^\intercal(\kappa_0+N_p-1)]^\intercal$ represents the continuous decision variables over the prediction horizon, ${\bm{\epsd}}(\kappa_0) = [{\epsd}^\intercal (\kappa_0), {\epsd}^\intercal(\kappa_0+1), \ldots, {\epsd}^\intercal(\kappa_0+N_p-1)]^\intercal$ expresses the discrete decision variables over the prediction horizon, and $N_p$ is the prediction horizon. Particularly, $\bm{\epsd}(\kappa_0)$ represents the train composition, and $\bm{\epsc}(\kappa_0)$ describes the departure and arrival times for each train service.
The system state is expressed as 
\begin{equation}\label{eq:augmented_state}
    \chi (\kappa) := [x^\intercal(\kappa_0),\ \bm{\rho}^\intercal(\kappa) ]^\intercal,
\end{equation}
where $\bm{x}(\kappa_0) = [{x}^\intercal (\kappa_0), {x}^\intercal(\kappa_0+1), \ldots, {x}^\intercal(\kappa_0+N_p-1)]^\intercal$ collects the dynamic component of the state trajectory over the prediction horizon, $\bm{\rho}(\kappa_0) = [\rho(\kappa_0)^\intercal, \rho(\kappa_0+1)^\intercal,\ldots, \rho(\kappa_0+N_\mathrm{p}-1)^\intercal]$ represents a concatenated vector with the expected passenger flows for each platform of the train line over the prediction horizon. 
More specifically, $x(\kappa_0)$ is composed of the number of passengers at each platform, the train composition for each running train service, and the number of trains in each depot of the line.
The optimization program \eqref{eq:minlp} can be either a mixed-integer nonlinear program (MINLP) or a linear mixed program (MILP) depending on whether the accurate nonlinear cost \eqref{eq:cost_passenger_nonlinear} or the approximate linear cost \eqref{eq:cost_passenger_linear} is chosen.

\section{State Reduction Approach for Learning-based MPC}\label{method}

This section focuses on the proposed state reduction approach, which integrates into the learning-based MPC framework presented in \cite{liu2025learningbasedmodelpredictivecontrol}. 
In the following, we describe the decoupling of discrete and continuous decision variables, the design of the state representation with different resolutions for learning and MPC, and the training and inference of the learning-based MPC approach with state reduction. 


\subsection{Decoupling of the decision variables}\label{sec:decoupling}

Consider a supervised learning classifier $\phi :\chi \mapsto \bm{\epsd}$ that approximates the mapping from the state $\chi$ into the optimal discrete decision variables $\bm{\epsd}^*$ over the prediction horizon, i.e., $\phi(\chi) \approx \bm{\epsd}^*(\chi)$,
where $\bm{\epsd}^*(\chi)$ is the solution of \eqref{eq:minlp} in the conditions specified by the state $\chi$.
During online operation, the predicted discrete variables $\bm{\epsd} = \phi(\chi)$ can be used to turn the original problem \eqref{eq:minlp} into a nonlinear (NLP) or linear (LP) problem, depending again on the choice of the performance index.
Then the continuous decision variables -- the departure and arrival times -- can be determined by the solution of the following optimization problem:
\begin{subequations}\label{eq:nlp}
\begin{align}
&{\mathop {\min }\limits_{\scriptstyle{\bm{x}}(\kappa_0),{\bm{\epsc}}(\kappa_0)} \!  J_\mathrm{d}^{(\mathrm{MPC})}(\chi(\kappa_0),\ \bm{\epsd}) := \sum\limits_{\kappa = {\kappa_0}}^{{\kappa_0}+ {N_p} - 1} {L^{(\epsd)}(x(\kappa),\epsc(\kappa))} } \label{eq:nlp_obj}\\
&\quad {\rm{s}}.{\rm{t}}.\quad {x}(\kappa + 1) = {A_\kappa}{x}(\kappa) + {B_{1,\kappa}}{\epsc}(\kappa) + {B_{2,\kappa}^{(\epsd)}},\label{eq:nlp_state_dynamics}\\
& \quad \qquad {D_{3,\kappa}}{x}(\kappa) + {D_{1,\kappa}}{\epsc}(\kappa)  \le  {D_{4,\kappa}^{(\epsd)}},\label{eq:nlp_constraints}\\
& \quad  \qquad \kappa = {\kappa_0}, \cdots ,{\kappa_0} + N_p - 1, \nonumber
\end{align}
\end{subequations}
where the cost $L^{(\epsd)}$ and the matrices (${B_{2,\kappa}^{(\epsd)}},{D_{4,\kappa}^{(\epsd)}}$) are introduced to reflect that $\epsd$ no longer is a decision variable.



\subsection{State reduction}

State dimensionality reduction in neural networks has several benefits, from training to inference. In a simpler state space, the neural network typically generalizes better, resulting in improved accuracy \citep{chandrashekar2014survey}. During training, the usage of computational resources can be lowered, as a lower-dimensional dataset occupies less memory. 
Alternatively, while maintaining the storage requirement identical, the number of points in the training set can be increased for a lower-dimensional state, which may enhance accuracy. Moreover, a lower-dimensional state space allows for a simpler neural network architecture since fewer neurons are required in each layer, allowing faster inference times. 
These benefits highlight the importance of proper design of the state space for learning. 
In what follows, we discuss a state reduction method that is tailored to the train rescheduling problem.

The core idea is to partition the vector $\bm{\rho}(\kappa)$ into a predetermined number $N_\mathrm{s}$ of segments, and then compute the average passenger flow for each segment. To illustrate this process, we consider the passenger flow at platform $p$ over a number $M=H\cdot N_\mathrm{s}$ of train services $\bm{\rho}_p=[\rho_p(k_p),...,\rho_p(k_p+M-1)]$, where $H$ is the length of the segments. Let $[\bm{\rho}_p]_j$ be the $j$th element of the vector $\bm{\rho}_p$ and $[\bm{\rho}_p]_{i:j}$ a slice of the vector $\bm{\rho}_p$ between the indices $i$ and $j$ provided that $i<j$. The state reduction procedure can be expressed as
\begin{equation}\label{eq:state_reduction}
\begin{split}
    [\bm{\rho}^\mathrm{reduced}_p]_{k}= \mathrm{mean(}[\bm{\rho}_p]_{k\cdot H:(k+1)\cdot H})
    \text{ for }k=0,...,N_\mathrm{s}
\end{split}
\end{equation}
where
$
    \mathrm{mean}([\bm{\rho}_p]_{i:j}) = \frac{1}{j-i}\sum_{q=i}^{j-1}[\bm{\rho}_p]_q
$.
This operation reduces the number of elements in the original vector by a factor of $H$. 
The state reduction preserves the mean of the original vector, ensuring no error in the computation of total 
Hereafter, the collection of the passenger flow $\bm{\rho}_p^\mathrm{reduced}$ across all platforms of a train line is denoted by $\bm{\rho}^\mathrm{reduced}$.
number of arriving passengers for each segment.

The discrete decision variable $\bm{\epsd}$, which represents the train composition of the train services over the prediction horizon, is relatively insensitive to fast changes in passenger flow over time. 
In this paper, we consider that train compositions can only be modified at depots located at the terminal stations of the line.  Thus, once the composition of the train of a particular train service is established, it has a lasting effect on the optimization problem \eqref{eq:nlp} as it cannot be changed on any of the intermediate platforms. 
Consequently, the variable $\bm{\epsd}$ is more sensitive to long temporal patterns of the passenger flow rather than short ones.
Therefore, we argue that a reduced passenger flow vector is more suitable for learning due to its lower dimensionality and its capacity to adequately capture the essential characteristics of the passenger flow required to estimate the optimal train composition.
Consequently, a compact state such as $\bm{\rho}_p^\mathrm{reduced}$, which has lower resolution than $\bm{\rho}_p$, may be used for the estimator $\phi$. Finally, the state described in \eqref{eq:augmented_state} can be rewritten as
\begin{equation}\label{eq:reduced_state}
    \chi^\mathrm{learning}(\kappa) = [x^\intercal(\kappa),\  (\bm{\rho}^{\mathrm{reduced}}(\kappa))^\intercal ]^\intercal,
\end{equation}
and the estimator $\phi$ uses this reduced state.

On the other hand, in the optimization problem \eqref{eq:nlp}, the unreduced passenger flow $\bm{\rho}$ is required to accurately model system dynamics. Furthermore, the continuous decision variable $\bm{\epsd}$, including departure and arrival times, has a relatively faster effect on \eqref{eq:nlp} and can react to faster changes in passenger flows. Hence, passenger flow $\bm{\rho}$ with the original resolution is more appropriate to be included in the MPC formulation of \eqref{eq:nlp} that computes $\bm{\epsd}$.

In essence, we advocate for the use of states with different levels of resolution for learning and optimization, leveraging the fact that the decision variables exhibit different sensitivities to changes in passenger flow. A low-resolution state is employed in learning to determine the discrete actions, while the original state representation is used in the MPC formulation to compute the continuous actions. In this manner, the state representation is tailored to the specific requirements of each component of the framework.

\subsection{Training and inference}

The procedure for the construction of the training dataset is described in Alg. \ref{alg:data_acquisition}. The training dataset is composed of tuples of states and their optimal solution for the discrete decision variables $\mathcal{D} = \{(\chi_\mathrm{learning}^{(k)},\ \bm{\epsd}^{(*,k)})\}_{k=1}^{N_\mathcal{D}}$. To generate a set of states that represent real operation, a number of episodes are started from random states within the operational range, and then the closed-loop trajectory with the optimal solution \eqref{eq:nlp} is simulated for a predetermined number of steps.

\begin{algorithm}[htb]
    \caption{Data acquisition}
    \label{alg:data_acquisition}
    \begin{algorithmic}[1] 
            \State \textbf{Input:} training dataset $\mathcal{D} \leftarrow \{\}$
            \For{episode = $1, ...,N_\mathrm{episodes}$}
                \State Set random state $\chi$
                \For{step = $1,...,N_\mathrm{steps}$}
                \State Get $\bm{\epsd}^*$ and $\bm{\epsd}^*$ by solving \eqref{eq:minlp} for state $\chi$
                \State Get $\chi_\mathrm{learning}$ via state reduction
                \State $\mathcal{D} \leftarrow \mathcal{D} \  \cup \  \{(\chi_\mathrm{learning},\ \bm{\epsd}^*)\}$
                \State Update $\chi$ by applying \eqref{eq:minlp_state_dynamics} with ($\bm{\epsd}^*,\ \bm{\epsd}^*$) 
                \EndFor
            \EndFor
            \State\textbf{Output:} training set $\mathcal{D}$
    \end{algorithmic}
\end{algorithm}

With the training set $\mathcal{D}$, the supervised learning classifier in $\phi$ can then be trained. 
We consider a set of hyperparameter configurations
$\mathcal{P}=\{\mathcal{P}_i\}_{i=1}^{n_\mathrm{NN}}$, such that for each $\mathcal{P}_i$, a classifier $\phi_i(\cdot)$ is trained via gradient descent.
Finally, the trained neural networks are stored in a set 
$\mathcal{N}=\{\phi_i\}_{i=1}^{n_\mathrm{NN}}$. 
As the choice of the classifier, we propose the use of recurrent neural networks, which are capable of encoding temporal dependencies in the discrete decision variables through their internal state.
In particular, LSTM networks are selected for their ability to represent long-term dependencies in sequential data.

The learning-based MPC approach becomes infeasible when the approximator fails to provide a discrete decision variable $\bm{\epsd}=\phi_i(\chi_\mathrm{learning})$ that renders the problem \eqref{eq:nlp} feasible. To mitigate this issue, we employ an ensemble of neural networks trained with heterogeneous hyperparameter configurations. Typically, after hyperparameter optimization, only the most accurate neural network is retained, while the others are discarded. In contrast, our approach evaluates each of the neural networks through closed-loop simulations of the system, as shown in Alg. \ref{alg:ensemble}. Based on this evaluation, a set $\mathcal{I} \subseteq \{1,...,n_\mathrm{NN}\}$ of the indices of the best-performing neural networks is selected and sorted from lowest to highest closed-loop total cost. 

\begin{algorithm}[htb]
    \caption{Formation of the ensemble}
    \label{alg:ensemble}
    \begin{algorithmic}[1] 
            \State \textbf{Input: } set of trained neural networks $\mathcal{N}$
            \For{$i=1,...,n_\mathrm{NN}$}
                \For{episode = $1, ...,N_\mathrm{test}$}
                \Comment{\textit{Closed-loop test}}
                    \State Set random state $\chi$
                    \For{step = $1,...,N_\mathrm{steps}$}
                    \State Get $\chi_\mathrm{learning}$ via state reduction
                    \State $\bm{\epsd} \leftarrow \phi_i(\chi_\mathrm{learning})$
                    \If{\eqref{eq:nlp} is feasible for $(\chi,\ \bm{\epsd})$}
                        \State Retrieve $\bm{\epsd}$
                    \Else{}
                        \State Use heuristics to find a feasible $\bm{\epsd}$
                        \State Get $\bm{\epsd}$ by solving \eqref{eq:nlp}
                    \EndIf
                    \State Update $\chi$ by applying \eqref{eq:nlp_state_dynamics} with ($\bm{\epsd},\ \bm{\epsd}$)
                    \EndFor
                \EndFor
            \EndFor
        \State \textbf{Output: } $\mathcal{I} \subseteq \{1,...n_\mathrm{NN}\}$: ordered set of indices of the networks arranged from lowest to highest total cost. 
    \end{algorithmic}
\end{algorithm}

\begin{figure}[tb]
    \centering
    \includegraphics[width=0.44\textwidth]{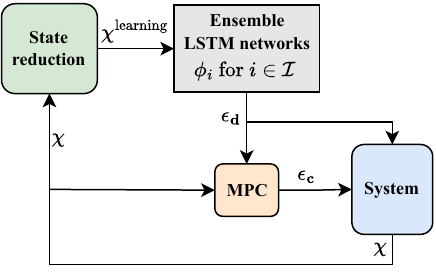}
    \caption{A representation of the control framework with state reduction.}
    \label{fig:control_framework}
\end{figure}

The online inference procedure is described in Alg. \ref{alg:online} and Fib. The core idea is to sequentially evaluate the trained neural networks in the ensemble, i.e., $\phi_i(\chi_\mathrm{learning})$ for $i \in \mathcal{I}$, until a feasible solution for the problem \eqref{eq:nlp} is found. If a feasible solution is not found by the supervised learning classifier, then heuristics described in can be used to restore feasibility, e.g., maintaining the same train composition of the previous time step or selecting the minimum number of train units for every train service, see \cite{liu2025learningbasedmodelpredictivecontrol} for more details.
By construction, feasibility also implies that all the constraints are satisfied and that the continuous decision variables $\bm{\epsd}$ are optimal with respect to the choice of $\bm{\epsd}$. As a result, integrating a learning-based approach with MPC can combine fast online evaluation with constraint satisfaction and optimality.
Although the neural network ensemble is evaluated sequentially due to its lower computational cost, the proposed approach is flexible to other types of evaluation, e.g., parallel evaluation followed by selection of the most common action or the one with the lowest cost.

\begin{algorithm}[htb]
    \caption{Online Inference}
    \label{alg:online}
    \begin{algorithmic}[1] 
            \State \textbf{Input:} set of indices $\mathcal{I}$, system state $\chi(\kappa)$
            \State $\chi \leftarrow \chi(\kappa)$
            \State $\texttt{solution\_found} \leftarrow False$
            \State Get $\chi_\mathrm{learning}$ via state reduction
            \For{$i \in \mathcal{I}$}
                \State $\bm{\epsd} \leftarrow \phi_i(\chi_\mathrm{learning})$
                \If{\eqref{eq:nlp} is feasible for $(\chi,\ \bm{\epsd})$}
                    \State Retrieve $\bm{\epsd}$
                    \State $\texttt{solution\_found} \leftarrow True$
                    \State Break for loop
                \EndIf
            \EndFor
            \If{$\texttt{solution\_found == False}$}
                \State Use heuristics to find a feasible $\bm{\epsd}$
                \State Get $\bm{\epsd}$ by solving \eqref{eq:nlp}
            \EndIf
            \State \textbf{Output: $\bm{\epsd}, \ \bm{\epsd}$} 
    \end{algorithmic}
\end{algorithm}


\section{Case study}\label{case_study}

\begin{figure}[htb]
\begin{center}
\includegraphics[width=0.48\textwidth]{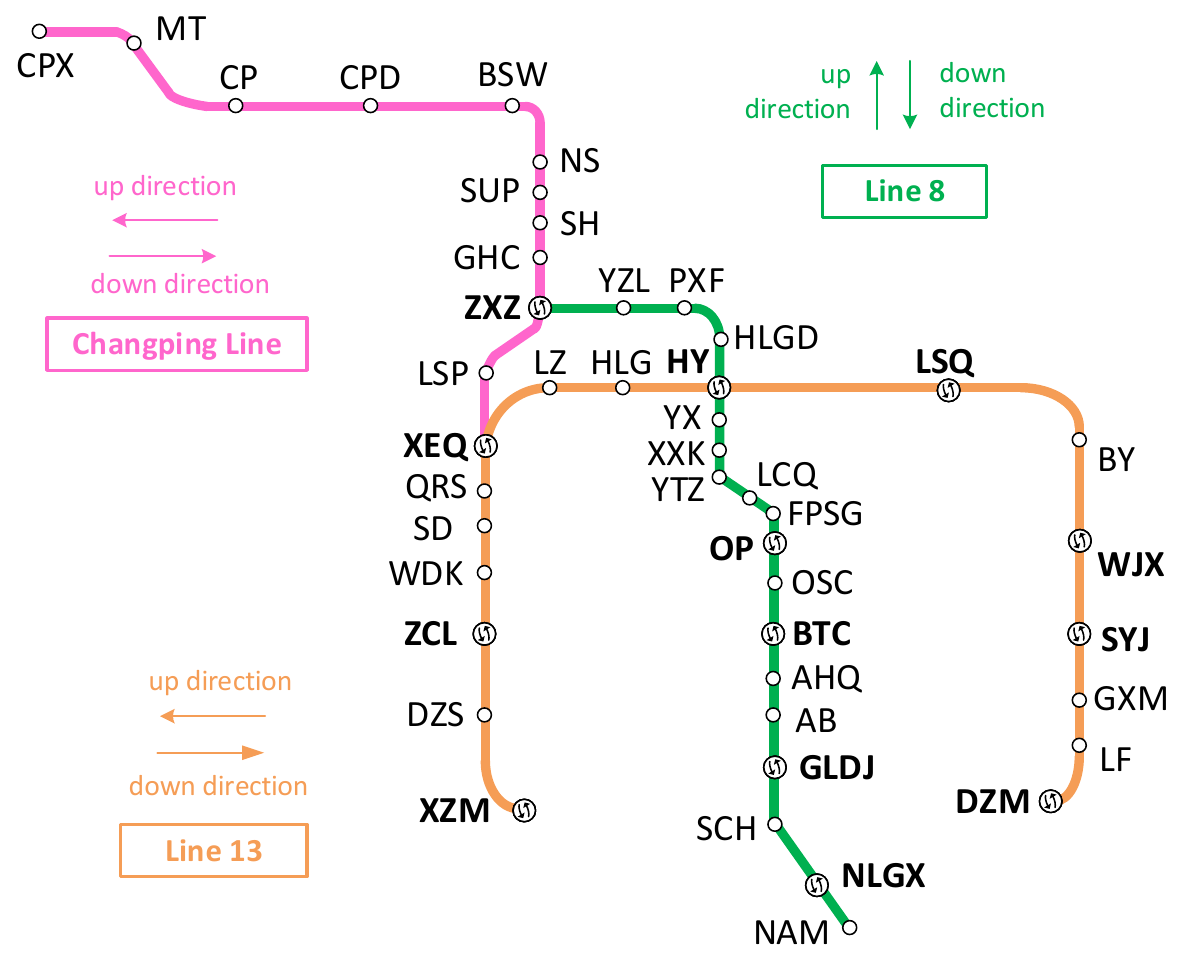}    
\caption{Three-line railway network in Beijing.} 
\label{fig:railnetwork}
\end{center}
\end{figure}

A three-line railway network in Beijing, shown in Fig. \ref{fig:railnetwork}, is used to assess the performance of the proposed approach. The network has 3 bidirectional lines, 45 stations, and 3 transfer stations (ZXZ, XEQ, HY). There is a depot connected to each line: CPX for Changping Line, XZM for Line 13, and ZXZ for Line 8. Moreover, the passenger flow data are based on real-world data.
The full description of the case study is provided in
\cite{liu2025learningbasedmodelpredictivecontrol}.
The experiments were conducted on Python with Intel XEON E5-6248R CPUs using PyTorch and Gurobi for machine learning and optimization, respectively.

The dataset characteristics for the full and reduced states, obtained by Alg. \ref{alg:data_acquisition}, are represented in Table \ref{table:dataset}. In our case, we had the two originally contrasting goals of (i) improving accuracy by enlarging the dataset and (ii) reducing memory requirements, which was one of the computational bottlenecks in our setup. Both goals were achieved by applying the state reduced approach \eqref{eq:state_reduction} with $N_\mathrm{s}=4$ and $H=9$. Consequently, the state dimension was reduced from $4002$ to $748$. Moreover, the number of data points could be increased by a factor of $2.65$, while decreasing the memory usage by a factor of $1.5$.
The datasets were then used for the training of $|\mathcal{P}|=64$ LSTM networks. Subsequently, a subset of $|\mathcal{I}|=15$ networks was selected to form the neural network ensemble according to Alg. \ref{alg:ensemble}.
Each hyperparameter configuration $\{\mathcal{P}_i\}_{i=1}^{64}$ included distinct values for the learning rate, number of hidden neurons, dropout rate, and two Boolean variables indicating whether output masking, and learning rate scheduling were applied.

\begin{table}[htb]
\begin{center}
\caption{Dataset information.}\label{table:dataset}
\resizebox{.48\textwidth}{!}{%
\begin{tabular}{cccc}
State & State dimension & Number of points & Memory (GB) \\\hline
Full & 4002 & 94871 & 5.00 \\
Reduced & 748 & 252207 & 3.29 \\ \hline
\end{tabular}
}
\end{center}
\end{table}

The comparison of the closed-loop operation of seven different methods over $4745$ samples is shown in Table \ref{tab:results}. The control time step and the MPC prediction horizon were set to $T = 4$ minutes and $N_p = 40$, respectively. Each closed-loop simulation was initialized at the same random state and had a maximum length of $30$ time steps. In methods (I)--(III), both the discrete and continuous variables are determined by optimization. The benchmark (I) corresponds to the solution of \eqref{eq:minlp} with a maximum solution time of $10$ minutes, rendering the method unachievable since the MPC controller must find a solution within the control time step.
Hence, benchmark (I) is used solely to define the optimality gap, measured as the percentage difference between the performance of a given method and that of benchmark (I).
The methods (II) and (III) represent the solution of \eqref{eq:minlp} with a maximum solution time of $4$ minutes -- equal to the control time step -- and nonlinear and linear performance indices, respectively.  In contrast, methods (IV)--(VII) employ the decoupling procedure described in Section \ref{sec:decoupling}, where the discrete variables are determined by an ensemble of neural networks, as described by Alg. \ref{alg:online}, and the continuous variables are computed by the solution of the optimization problem \eqref{eq:nlp} with a maximum solution time of $4$ minutes. 
The approaches (IV) and (V) use the original state representation, while (VI) and (VII) employ the reduced state described in \eqref{eq:reduced_state}. Although methods (V) and (VII) use the linear performance index \eqref{eq:cost_passenger_linear}, their solutions were evaluated on the nonlinear index \eqref{eq:cost_passenger_nonlinear}, ensuring consistency in the evaluation of all the methods. 

As illustrated by Table \ref{tab:results}, the optimality gaps of the learning-based approaches with unreduced state (IV) and (V) are significantly reduced by the proposed approaches (VI) and (VII).
More specifically, comparing (IV) to (VI) the optimality gap dropped from $7.38\%$ to $0.25\%$, and comparing (V) to (VII) the optimality gap decreases from $8.03\%$ to $0.15\%$, showing the significant effect of the state reduction procedure \eqref{eq:state_reduction}.
The learning-based approaches (IV)--(VII) reduce the computation time considerably compared to the optimization-based approaches (I)--(III).
Particularly, the learning-based approach (VII) with the linear performance index \eqref{eq:cost_passenger_linear} has the best overall performance considering optimality and computation time.
In methods (IV) and (VI), the NLP \eqref{eq:nlp} is not solved to optimality, as an early termination criterion is employed to reduce the CPU time. 
This heuristic considerably reduces the computation at the expense of a minimal increase in the optimality gap. This accounts for the slightly larger optimality gap of (VI) when compared to that of (VII).


\begin{table}[tb]
\centering
\caption{Comparison of the closed-loop performance for several approaches, with the proposed approach highlighted in bold.}
\label{tab:results}
\resizebox{.48\textwidth}{!}{%
\begin{tabular}{c|ccc|cll}
\hline
\multirow{2}{*}{Method} & \multicolumn{3}{c|}{Optimality gap (\%)} & \multicolumn{3}{c}{CPU time (s)} \\ \cline{2-7} 
 & mean & max & std & mean & max & std \\ \hline
(I) Benchmark & 0.00 & 0.00 & 0.00 & 196.00 & 600.40 & 272.50 \\ \cline{1-1}
(II) MINLP & 0.00 & 3.07 & 0.24 & 26.99 & 240.1 & 39.01 \\ \cline{1-1}
(III) MILP & 0.30 & 1.47 & 0.35 & 4.00 & 115.3 & 10.13 \\ \hline
(IV) Learning + NLP & 7.38 & 44.66 & 9.61 & 6.88 & 56.07 & 7.18 \\ \cline{1-1}
(V) Learning + LP & 8.03 & 44.27 & 10.27 & 0.15 & 0.37 & 0.11 \\ \hline
\textbf{\begin{tabular}[c]{@{}c@{}}(VI) Learning + NLP\\ with state reduction\end{tabular}} & \textbf{0.25} & \textbf{5.58} & \textbf{1.30} & \textbf{6.65} & \textbf{30.45} & \textbf{5.62} \\ \cline{1-1}
\textbf{\begin{tabular}[c]{@{}c@{}}(VII) Learning + LP\\ with state reduction\end{tabular}} & \textbf{0.15} & \textbf{3.61} & \textbf{0.92} & \textbf{0.16} & \textbf{0.41} & \textbf{0.03} \\ \hline
\end{tabular}%
}
\end{table}



\section{Conclusions}\label{conclusion}

This paper has proposed a state reduction scheme for learning-based MPC in the context of train rescheduling, where different levels of resolution are used to represent the system state for learning and control.
Simulation experiments conducted on a large-scale railway network revealed a dramatic decrease in suboptimality when the proposed state reduction was employed, thereby confirming its effectiveness.
Future research will explore alternative state reduction techniques such as principal component analysis and autoencoders.



\begin{ack}
This research has received funding from the European Research Council (ERC) under the European Union’s Horizon 2020 research and innovation programme (Grant agreement No. 101018826 - CLariNet).
\end{ack}

\bibliography{ifacconf}            


\end{document}